\documentclass[pra,twocolumn,showpacs]{revtex4}
\usepackage{fancyhdr}
\usepackage{amssymb}
\usepackage{amsmath}
\usepackage[mathscr]{eucal}
\usepackage{latexsym}
\usepackage{amsbsy}
\usepackage{float}
\usepackage[dvips]{graphicx}
\usepackage{epsfig}
\usepackage{subfigure}
\usepackage{wrapfig}

\newcommand{\be}{\begin{equation}}
\newcommand{\ee}{\end{equation}}
\newcommand{\bes}{\begin{equation*}}
\newcommand{\ees}{\end{equation*}}
\newcommand{\bea}{\begin{eqnarray}}
\newcommand{\eea}{\end{eqnarray}}
\newcommand{\bean}{\begin{eqnarray*}}
\newcommand{\eean}{\end{eqnarray*}}
\newcommand{\ba}{\begin{array}}
\newcommand{\ea}{\end{array}}

\newcommand{\Tr}{\mathrm{Tr}}
\newcommand{\vk}{{\bf k}}
\newcommand{\vq}{{\bf q}}
\begin{document}

\title{\large Condensate fraction of a resonant Fermi gas
with spin-orbit coupling\\ in three and two dimensions}

\author{L. Dell'Anna, G. Mazzarella, and L. Salasnich}

\affiliation{Dipartimento di Fisica ``Galileo Galilei'' and CNISM,
Universit\`a di Padova, I-35122 Padova, Italy}

\begin{abstract}
We study the effects of laser-induced Rashba-like spin-orbit coupling along
the Bardeen-Cooper-Schrieffer--Bose-Einstein condensate (BCS-BEC) crossover of 
a Feshbach resonance for a two-spin-component
Fermi gas. We calculate the condensate fraction in three and two
dimensions and find that this quantity characterize the crossover
better than other quantities, like the chemical potential
or the pairing gap. By considering both the singlet and the triplet pairings, 
we calculate the condensate fraction and show that a large enough
spin-orbit interaction enhances the singlet condensate fraction in the 
BCS side while suppressing it on the BEC side.
\end{abstract}


\maketitle

\section{Introduction}

Over the past several years, 
the predicted crossover \cite{eagles,leggett,noziers}
from the Bardeen-Cooper-Schrieffer (BCS)
state of weakly bound Fermi pairs to the Bose-Einstein condensate (BEC)
of molecular dimers has been observed by several experimental groups
\cite{greiner,regal,kinast,zwierlein,chin,ueda}.
In two experiments \cite{zwierlein,ueda} the condensate fraction
of Cooper pairs \cite{yang}, which is directly related
to the off-diagonal-long-range order of the two-body density
matrix of fermions \cite{penrose,campbell}, has been studied with
two hyperfine component Fermi vapors of $^6$Li atoms in
the BCS-BEC crossover. The experimental data are in quite good agreement
with mean-field theoretical predictions
\cite{sala-odlro,ortiz} and Monte-Carlo
simulations \cite{astrakharchik} at zero temperature,
while at finite temperature beyond-mean-field
corrections are needed \cite{ohashi}.
Recently the condensate fraction in the BCS-BEC crossover for a
two-dimensional (2D) Fermi gas \cite{sala-odlro2}, and for a
three-spin-component Fermi gas with SU(3) symmetry \cite{sala-odlro3},
has been theoretically investigated. Remarkably,
last year 2D degenerate Fermi gases were experimentally realized
for ultra-cold atoms in a highly anisotropic
disk-shaped potential \cite{turlapov}.

Quite recently, artificial spin-orbit coupling has been obtained
in neutral bosonic systems \cite{spielman}, where the strength of the
coupling can be controlled optically, and it has been suggested
that the same techniques can be used with ultracold
fermions \cite{dalibard,chapman}. These results have stimulated
the theoretical investigation of spin-orbit effects
with Rashba \cite{rashba} and Dresselhaus \cite{dress} terms
in the BCS-BEC crossover \cite{zhang,zhai,hu,sa,cin,andrea}.
In particular, very recently and
independently,
several authors have analyzed the evolution
from BCS to BEC superfluidity in the presence of spin-orbit coupling
for a 3D uniform Fermi gas \cite{zhang,zhai,hu,sa},
and in the 2D case
by use of a perturbative approach \cite{cin}.
Nevertheless, those papers
did not considered the condensate fraction of Fermi pairs.

In the present paper we calculate the chemical potential, the pairing gap,
and the condensate fraction along the BCS-BEC crossover both in 3D and 2D
as a function of spin-orbit coupling.
We show that the two contributions - i.e., those related to the singlet 
and triplet pairings - to the condensate fraction, 
separately, characterize the crossover
better than the other quantities. Remarkably, 
a large enough spin-orbit interaction enhances 
the singlet contribution to the condensate
fraction in the BCS side while suppressing it on the BEC side. The triplet 
contribution to the condensate grows by increasing the spin-orbit coupling and 
is larger close to the crossover. On the contrary,
the chemical potential and the pair function exhibit no peculiarities
along the crossover. Moreover, we find that, when the Rashba velocity becomes
of the order of the Fermi velocity, there is a value for the dimensionless
interaction strength $y=1/(k_Fa_s)$, where $k_F$ denotes 
the Fermi linear momentum
and $a_s$ the inter-atomic $s$-wave scattering length,
for which the singlet condensate fraction no longer depends on 
spin-orbit coupling.
This nodal point can be promoted as the real point of the crossover.
What is observed in three dimensions occurs also in two dimensions where the
nodal point occurs when the binding energy $\epsilon_B$
is almost equal to the Fermi energy $\epsilon_F$,
i.e., $\epsilon_B\approx \epsilon_F$. In 2D
the condensate fraction approaches the value $1$
only for extremely large values of the scaled
binding energy $\epsilon_B/\epsilon_F$.

\section{The model}

Let us consider the following Hamiltonian
\be
H=H_0+H_I \; ,
\ee
where $H_0$ is the single particle Hamiltonian in the presence of Rashba
and Dresselhaus terms \cite{dress,rashba}, namely
\bea
H_0&=&\sum_{\vk}\psi(\vk)^\dagger\Big\{\frac{\hbar^2 k^2}{2m}
+\hbar\big[v_R(\sigma_x k_y-\sigma_y k_x)
\nonumber
\\
&+&v_D(\sigma_x k_y+\sigma_y k_x) \big]\Big\}\psi(\vk)\;.
\eea
where $v_D$ and $v_R$ are, respectively, the Rashba and Dresselhaus velocities;
$\sigma_x$ and $\sigma_y$ denote the Pauli matrices in the $x$ and $y$
directions; and $\psi(\vk)$ is the spinor
$\psi(\vk)=(\psi_{\uparrow}(\vk),\psi_{\downarrow}(\vk))^T$.
$H_I$ is the interaction term given by
\be
H_I=-\frac{g}{V} \sum_{\vk \vk' \vq}\psi_{\uparrow}^\dagger(\vk+\vq)\psi_{\downarrow}^\dagger(-\vk)
\psi_{\downarrow}(-\vk'+\vq)\psi_{\uparrow}(\vk') \;,
\ee
where $g>0$, which corresponds to attractive interaction.
After defining the order parameter describing the particle pairs,
$\Delta=(g/V)\sum_\vk \langle
\psi_\downarrow(-\vk) \psi_\uparrow(\vk)\rangle$, where $V$ is the volume,
at the mean field level we can decouple the interaction, finding
\be
H_I=V\frac{|\Delta|^2}{g}-\sum_\vk\left(\Delta^*\psi_\downarrow(-\vk)\psi_\uparrow(\vk)+\Delta
\psi_\uparrow^\dagger(\vk) \psi_\downarrow^\dagger(-\vk)\right)\; .
\ee
Introducing the following multispinor $\Psi(\vk)=(\psi_\uparrow(\vk),
\psi_\downarrow^\dagger(-\vk),\psi_\downarrow(\vk),\psi_\uparrow^
\dagger(-\vk))^T$, one can resort to standard path integral formulation
at finite temperature obtaining, within a saddle point approximation and
after integrating over the fermions \cite{stoof}, the thermodynamic potential
\be
\Omega=V\frac{|\Delta|^2}{g}-\frac{1}{2\beta}\sum_{\vk\, \omega}\Tr
\ln G^{-1}+\sum_\vk\xi_{\vk} \; .
\ee
where $\beta = 1/(k_B T)$, where $k_B$ denotes the Boltzmann constant and $T$ 
the absolute temperature, and $G^{-1}$ is a matrix on the basis of
$\Psi(\vk)$ which reads
\be
G^{-1}(\vk,\omega)=\left(
\ba{cccc}
i\omega+\xi_\vk & -\Delta& \gamma(\vk)&0\\
-\Delta^*&i\omega-\xi_\vk & 0 &-\gamma(-\vk)\\
\gamma^*(\vk)&0&i\omega+\xi_{\vk}&\Delta\\
0&-\gamma^*(-\vk)&\Delta^*&i\omega-\xi_{\vk}
\ea
\right)
\ee
with $\gamma(\vk)=\hbar v_R(k_y+ik_x)+\hbar v_D(k_y-ik_x)$ and
$\xi_{\vk}=\hbar^2k^2/2m-\mu$. After summing over the Matsubara
frequencies \cite{stoof} the thermodynamic potential becomes
\be
\Omega=V\frac{|\Delta|^2}{g}-\frac{1}{2\beta}\sum_{\vk}\sum_{i=1}^4
\ln(1+e^{-\beta E_i(\vk)})+\sum_\vk\xi_{\vk} \; ,
\ee
where $E_1(\vk)=\sqrt{(\xi_\vk-|\gamma(\vk)|)^2+|\Delta|^2}$, $E_2(\vk)
=\sqrt{(\xi_\vk+|\gamma(\vk)|)^2+|\Delta|^2}$, $E_3=-E_1$ and $E_4=-E_2$.
From the thermodynamic formula
$N=-\frac{\partial \Omega}{\partial \mu}$ we obtain
the equation for the number of particles
\bea
N&=&\sum_{\vk}\Big\{1-\tanh\left(\beta E_1(\vk)/2\right)\frac{\xi_{\vk}-
|\gamma(\vk)|}{2E_1(\vk)}
\nonumber
\\
&-&\tanh\left(\beta E_2(\vk)/2\right)\frac{\xi_{\vk}+
|\gamma(\vk)|}{2E_2(\vk)}\Big\} \; .
\eea
The gap equation is, instead, given by
\be
\label{gap}
\frac{V}{g}=\frac{1}{4}\sum_\vk\left(\frac{\tanh\left(\beta E_1(\vk)/2\right)}
{E_1(\vk)}+\frac{\tanh\left(\beta E_2(\vk)/2\right)}{E_2(\vk)}\right) \; ,
\ee
and finally the condensate number \cite{leggett2,campbell,sala-odlro} reads
\be
N_C=N_{0}+N_{1},
\ee
where 
\bean
N_{0}&=&\sum_\vk \left|\langle \psi_\uparrow(\vk)\psi_\downarrow(-\vk)
\rangle\right|^2\nonumber
\\
&=& \frac{|\Delta|^2}{16}
\sum_\vk\left(\frac{\tanh\left(\beta E_1(\vk)/2\right)}
{E_1(\vk)}+\frac{\tanh\left(\beta E_2(\vk)/2\right)}{E_2(\vk)}\right)^2 \;
\eean
is the singlet contribution to the condensate, with total spin $0$, whereas
\bean
N_{1}&=&\sum_\vk \left|\langle \psi_\uparrow(\vk)\psi_\uparrow(-\vk)
\rangle\right|^2\nonumber
\\
&=& \frac{|\Delta|^2}{16}
\sum_\vk\left(\frac{\tanh\left(\beta E_1(\vk)/2\right)}
{E_1(\vk)}-\frac{\tanh\left(\beta E_2(\vk)/2\right)}{E_2(\vk)}\right)^2 \; 
\eean
is the triplet one, with total spin $1$.

\begin{figure}[ht]
\includegraphics[width=8.cm]{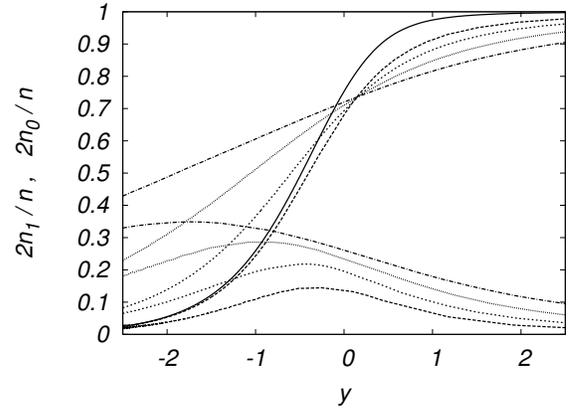}
\caption{Singlet ($n_{0}$, upper curves) and triplet ($n_{1}$, lower curves) 
condensate fractions of the 3D Fermi gas as functions of the dimensionless
interaction strength $y=1/(k_Fa_s)$ for different values
of Rashba velocity, $(v_R/v_F)^2=0$ (solid line), $0.5$ (long-dashed line),
$1$ (short-dashed line), $2$ (dotted line), and $4$ (dashed-dotted line).}
\label{fig.3d_f1}
\end{figure}

\begin{figure}[ht]
\includegraphics[width=6.5cm]{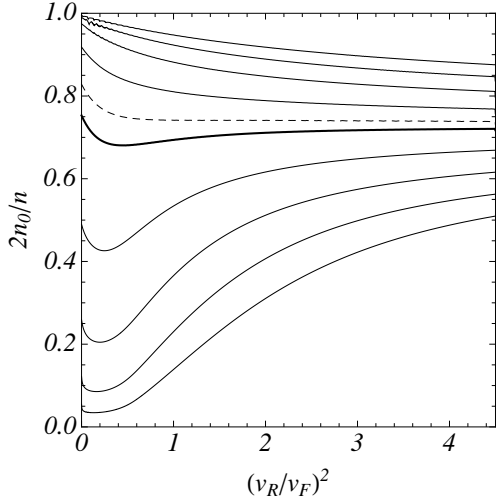}
\caption{Singlet condensate fraction of the 3D Fermi gas
as a function of $(v_R/v_F)^2$ for $y=-2,
-1.5, -1, -0.5, 0, 0.18, 0.5, 1, 1.5, 2$ (corresponding to the curves
from below). The solid thicker line is for $y=0$. The dashed line is
when $y\approx 0.2$, below this value all the curves have a minimum,
while above it they always decrease.}
\label{fig.3d_f2}
\end{figure}

We are interested in the low temperature regime where the condensate is quite
large. Quantitatively we can restrict our study to the zero temperature
limit, or, in three dimensions, when at least $2k_BT\ll \Delta$, with
$\Delta$ now supposed to be a real number.
The zero temperature limit is also mandatory for the two dimensional case.
In the equations above we have therefore simply $\tanh(\beta E_i(\vk)/2)
\rightarrow 1$.

\section{Three dimensions}

Let us consider first the three dimensional case. Hereafter
we proceed in the same spirit of Ref.~\cite{marini}, generalizing
the calculation including the spin-orbit coupling. After rescaling the momenta
\be
\label{resc}
\vk= \frac{\sqrt{2m\Delta}}{\hbar}\vq
\ee
and summing in the continuum
($\sum_{\vk}\rightarrow \frac{V}{(2\pi)^3}\int d^3{\vk}$) we get, for the
number of particles
\be
\label{N}
n=\frac{N}{V}=\frac{(2m\Delta)^{3/2}}{(2\pi\hbar)^3}\,
I^{3d}_{N}(x_0,x_1,x_2) \; ,
\ee
where
\bea
I^{3d}_{N}(x_0,x_1,x_2)=\int d^3{\vq}
\Big(1-\frac{1}{2}\sum_{r=\pm 1}
\nonumber
\\
\frac{q^2-x_0+r\sqrt{x_1^2 q_x^2+x_2^2 q_y^2}}{\sqrt{\left(q^2-x_0+
r\sqrt{x_1^2 q_x^2+x_2^2 q_y^2}\right)^2+1}}\Big)
\label{IN3d}
\eea
with dimensionless parameters defined as follows
\bea
&& x_0=\frac{\mu}{\Delta}\\
&& x_1=2m \frac{(v_R-v_D)^2}{\Delta}\\
&& x_2=2m \frac{(v_R+v_D)^2}{\Delta} \; .
\eea
In the continuum limit, due to the
choice of a contact potential, the gap equation (\ref{gap})
diverges in the ultraviolet. After regularization \cite{leggett}
the gap equation reads
\be
\frac{1}{g}=-\frac{m}{4\pi \hbar^2 a_s}+\frac{1}{V}\sum_\vk \frac{1}
{2(\xi_\vk+\mu)} \; ,
\ee
where $a_s$ is the $s$-wave scattering length between fermions with
different spin component. In this way we get
\be
\label{y}
y\equiv \frac{1}{k_F a_s}=\frac{1}{3^{1/3}\pi^{5/3}}\frac{I_{a_s}
(x_0,x_1,x_2)}{I^{3d}_N(x_0,x_1,x_2)^{1/3}} \; ,
\ee
where
\bea
I_{a_s}(x_0,x_1,x_2)=\int d^3{\vq} \Big(\frac{1}{q^2}-\frac{1}{2}\sum_{r=\pm 1}
\nonumber
\\
\frac{1}{\sqrt{\left(q^2-x_0+r\sqrt{x_1^2 q_x^2+x_2^2 q_y^2}\right)^2+1}}
\Big) \; .
\eea

\begin{figure}[ht]
\includegraphics[width=8.cm]{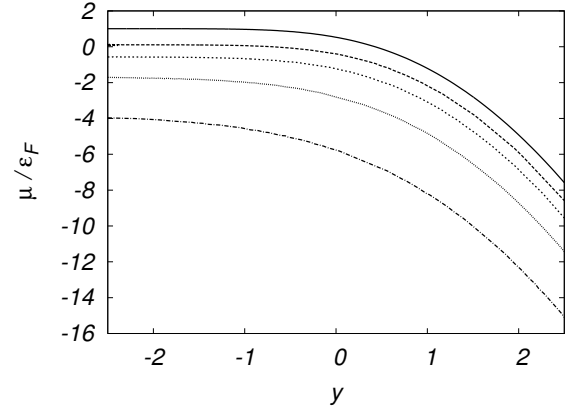}
\includegraphics[width=8.cm]{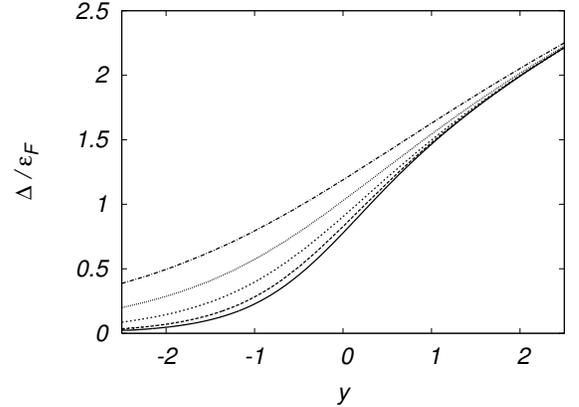}
\caption{3D Fermi gas: Chemical potential $\mu$ (upper panel)
and pair function $\Delta$ (lower panel), both in units of $\epsilon_F$,
as a function of the dimensionless interaction strength
$y=1/(k_Fa_s)$ for different values of $(v_R/v_F)^2$ 
(the same ones as in Fig.~\ref{fig.3d_f1}, with same line types).}
\label{fig.3d_all}
\end{figure}

Finally the condensate densities are given by
\be
n_{s}=\frac{N_{s}}{V}=\frac{(2m\Delta)^{3/2}}{16(2\pi\hbar)^3}I^{3d}_{N_{s}}
(x_0,x_1,x_2) \; ,
\ee
where $s=0,1$, and
\bea
I^{3d}_{N_{s}}(x_0,x_1,x_2)=\int d^3{\vq} \Big(\sum_{r=\pm 1}
\nonumber
\\
\frac{r^s}
{\sqrt{\left(q^2-x_0+r\sqrt{x_1^2 q_x^2+x_2^2
q_y^2}\right)^2+1}}\Big)^2 \; .
\label{IN03d}
\eea

We can write also the gap and the chemical potential in terms of the Fermi
energy $\epsilon_F=\hbar^2 k_F^2/(2m)=\hbar^2/(2m)(3\pi^2 n)^{2/3}$,
therefore
\bea
\label{gap_3d}
&&\frac{\Delta}{\epsilon_F}=4\left(\frac{\pi}{3}\right)^{2/3}
I^{3d}_{N}(x_0,x_1,x_2)^{-2/3}\; ,\\
&&\frac{\mu}{\epsilon_F}=4\left(\frac{\pi}{3}\right)^{2/3} x_0
I^{3d}_{N}(x_0,x_1,x_2)^{-2/3} \; .
\label{mu_3d}
\eea
Finally the spin-orbit velocities can be written in terms of the Fermi velocity
\be
\frac{(v_R\mp v_D)^2}{v_F^2}=\left(\frac{\pi}{3}\right)^{2/3}x_{1,2}\,
I^{3d}_N(x_0,x_1,x_2)^{-2/3}
\ee
We are now in the position to express the two contributions to the condensate fraction
\be
\frac{2n_{s}}{n}=\frac{1}{8}\frac{I^{3d}_{N_{s}}(x_0,x_1,x_2)}
{I^{3d}_N(x_0,x_1,x_2)},
\ee
the chemical potential, Eq.~(\ref{mu_3d}), and the gap, Eq.~(\ref{gap_3d}), in
terms of the scattering parameter $y$, Eq.~(\ref{y}). This is guaranteed, at
least heuristically, by the fact that for any point in the space of
dimensionless parameters, $(x_0,x_1,x_2)$, there are single values for $y$,
$2n_{s}/n$, $\mu/\epsilon_F$ and $\Delta/\epsilon_F$. For $x_1=x_2=0$, namely
without spin-orbit couplings, we indeed recover previous analytic results
reported in Ref.~\cite{marini}.

The results shown here in the figures are obtained fixing $x_1=x_2$,
namely when only Rashba ($v_D=0$) or only Dresselhaus ($v_R=0$) are present.
The other special case with $v_R=v_D$ is actually less interesting since in
that case the singlet condensate fraction seems to be always suppressed.
On the contrary, with only Rashba term (or only
Dresselhaus) we observe (see Fig.~\ref{fig.3d_f1}) that, turning on the
spin-orbit coupling, the singlet condensate fraction increases in the 
BCS regime, 
whereas it decreases in the BEC regime. In particular, for $v_R \gtrsim v_F$ 
the singlet condensate fraction at $y\approx 0.2$, slightly above the 
unitarity, which 
is $2n_0/n\approx 0.7$, does no longer depends on
the spin-orbit coupling; see Fig.~\ref{fig.3d_f2}. On the left (BCS side)
of this point the singlet condensation is improved by the spin-orbital 
interaction,
whereas on the right (BEC side) this condensation is suppressed. 
The triplet condensate fraction, instead, decreases in both the BCS and 
BEC limits, whereas it is sizable close to the crossover, exhibiting a non-monotonic behavior. 
On the contrary, the chemical potential $\mu$ (upper panel of
Fig.~\ref{fig.3d_all}) is shifted
toward negative values in both the regimes, while the pair function
$\Delta$ (lower panel of Fig.~\ref{fig.3d_all}) is enhanced both in the 
BCS side and in the BEC one, although,
in the latter, the enhancement is less pronounced. These last quantities,
therefore, unlike the condensate fraction, exhibit no 
peculiarities across the crossover.

\section{Two dimensions}

In two dimensions the regularization of the gap equation differs, with a bound 
state always present \cite{randeria}. With $\epsilon_B$ as the
binding energy, we have, therefore,
\be
\frac{1}{g}=\frac{1}{V}\sum_\vk\frac{1}{2(\xi_\vk+\mu)+\epsilon_B} \; ,
\ee
which, after rescaling the momenta as in Eq.~(\ref{resc}) and integrating,
leads to
\be
\frac{1}{g}=\frac{m}{4\pi\hbar^2}\ln\left(\frac{2\Lambda^2}{\epsilon_B/\Delta}
+1\right) \; ,
\ee
where $\Lambda$ is the ultraviolet momentum cut-off.
On the other hand Eq.~(\ref{gap}) holds, where now the sum is over momenta
in two dimensions, and, therefore, we have
\be
\frac{1}{g}=\frac{m}{(2\pi\hbar)^2}I_g(x_0,x_1,x_2)
\ee
with
\bea
I_g(x_0,x_1,x_2)=\frac{1}{2}\int^\Lambda d^2{\vq} \Big(\sum_{r=\pm 1}
\nonumber
\\
\frac{1}{\sqrt{\left(q^2-x_0+r\sqrt{x_1^2 q_x^2+x_2^2 q_y^2}\right)^2+1}}
\Big) \; .
\eea

\begin{figure}[ht]
\includegraphics[width=8.cm]{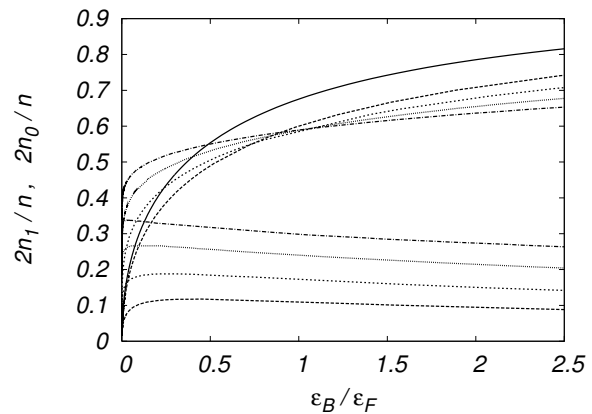}
\caption{Singlet ($n_{0}$, upper curves) and triplet ($n_{1}$, lower curves) condensate fractions of the 2D Fermi gas
as functions of the binding energy
$\epsilon_B$, in units of the Fermi energy $\epsilon_F$, for
different values of Rashba velocity, $(v_R/v_F)^2=0$ (solid line),
$0.5$ (long-dashed line), $1$ (short-dashed line), $2$ (dotted line),
and $4$ (dashed-dotted line).}
\label{fig.2d_f1}
\end{figure}

\begin{figure}[ht]
\includegraphics[width=6.5cm]{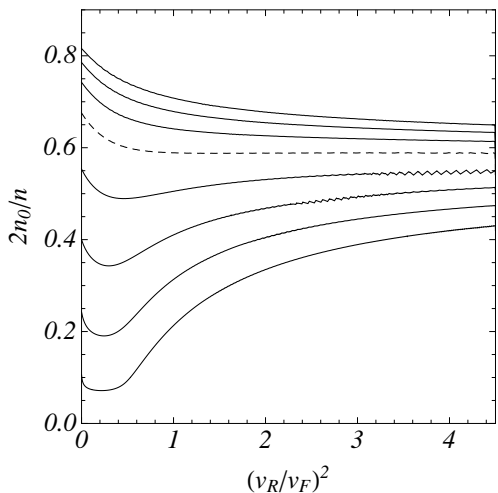}
\caption{Singlet condensate fraction of the 2D Fermi gas
as a function of $(v_R/v_F)^2$ for
$\epsilon_B/\epsilon_F=0.01, 0.06, 0.2, 0.5, 1, 1.5, 2, 2.5$
(corresponding to the curves from below). The dashed line is when
$\epsilon_B/\epsilon_F \approx 1$, below this value all the curves
have a minimum, whereas, above it, they always decrease.}
\label{fig.2d_f2}
\end{figure}

From this expression we derive the binding energy
\be
\frac{\epsilon_B}{\Delta}=\lim_{\Lambda\rightarrow \infty}
\frac{2\Lambda^2}{\exp[I_g(x_0,x_1,x_2)/\pi]-1} \; ,
\ee
which actually does not depend on the cutoff since $I_g$ has a
logarithmic divergence in the ultraviolet which cancels exactly the factor
$\Lambda^2$. In the absence of spin-orbit, $x_1=x_2=0$, we recover, in fact,
the known result $\epsilon_B/\Delta=\sqrt{x_0^2+1}-x_0$. \cite{marini}

As in the three dimensional case, the quantities we consider
are the following
\bea
&&\frac{2n_{s}}{n}=\frac{1}{8}\frac{I^{2d}_{N_{s}}(x_0,x_1,x_2)}
{I^{2d}_N(x_0,x_1,x_2)},\\
&&\frac{\mu}{\epsilon_F}=\frac{2\pi x_0}{I^{2d}_N(x_0,x_1,x_2)}\\
&&\frac{\Delta}{\epsilon_F}=\frac{2\pi}{I^{2d}_N(x_0,x_1,x_2)}
\eea
as functions of $\epsilon_B/\epsilon_F$ where now $\epsilon_F=\hbar^2 \pi n/m$
is the Fermi energy in two dimensions. The integrals $I^{2d}_N$ and
$I^{2d}_{N_{s}}$ are the same ones as in Eqs.~(\ref{IN3d}) and 
(\ref{IN03d}) but
defined in two dimensions ($d^3\vq \rightarrow d^2\vq$).

Experimentally, the realization of a 2D system can be obtained by a strong
harmonic confinement in one direction, i.e., $\omega_z\gg \omega_x, \omega_y$;
therefore one can link the tunable 3D
scattering length $a_s$ to the two-body binding energy in 2D.
Introducing, conveniently, 
the confining length $\ell_z=\sqrt{h/m\omega_z}$, one finds, in fact,
$\ln(\epsilon_B/\hbar\omega_z)\sim \ell_z/a_s$ (for more details see Refs.
\cite{morais, bertaina}).

Again we focus our attention to the case with only Rashba (or, equivalently,
only Dresselhaus) term, i.e., $x_1=x_2$.
\begin{figure}[ht]
\includegraphics[width=8.cm]{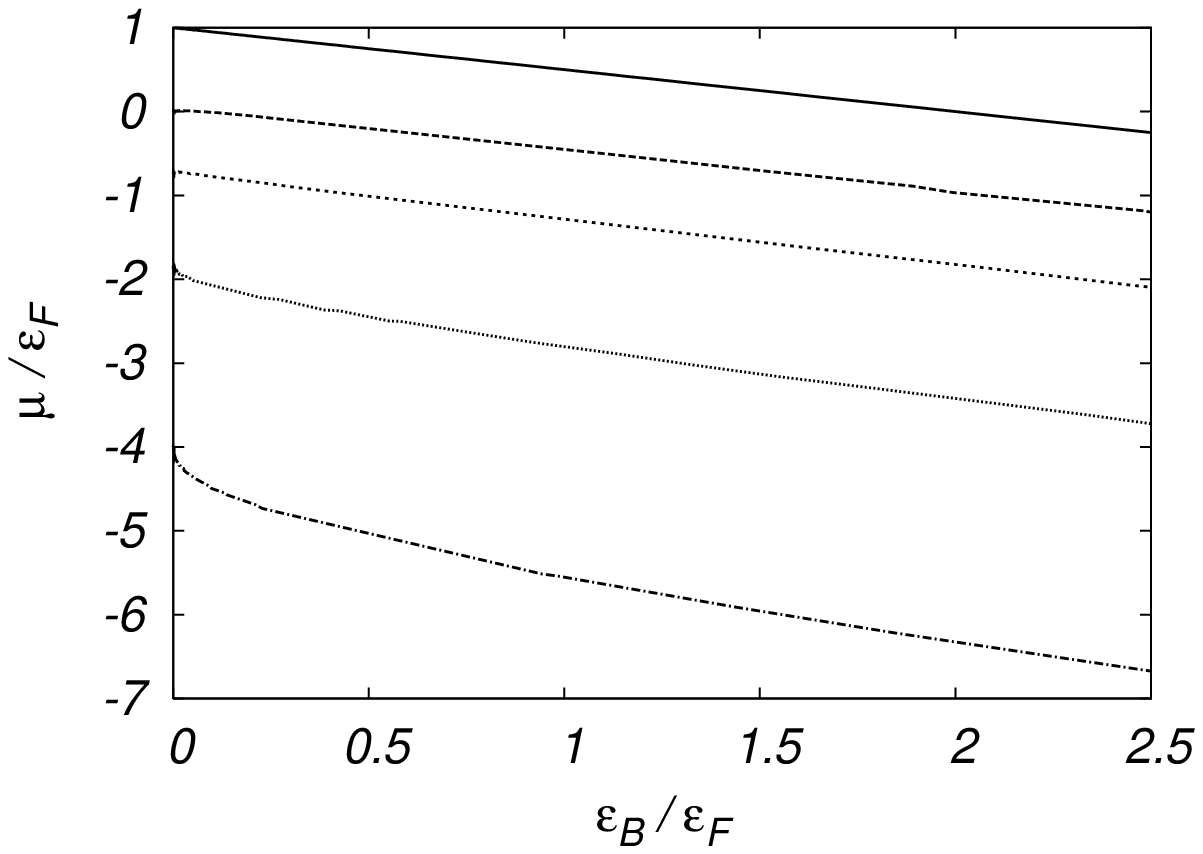}
\includegraphics[width=8.cm]{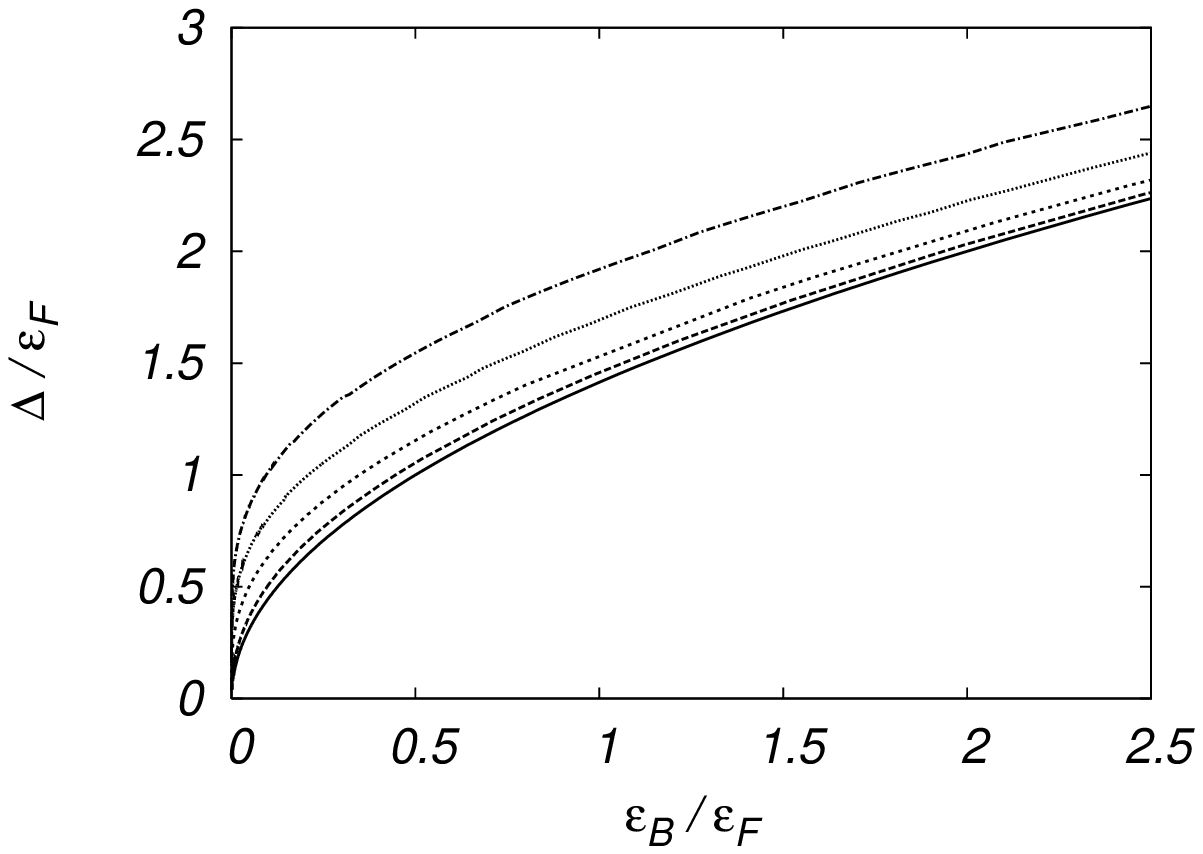}
\caption{2D Fermi gas: Chemical potential $\mu$ (upper panel)
and pair function $\Delta$ (lower panel), both
in units of $\epsilon_F$, as a function
of the binding energy $\epsilon_B$
(in units of $\epsilon_F$) for different values
of $(v_R/v_F)^2$ (the same ones as in Fig.~\ref{fig.2d_f1} 
with same line types).}
\label{fig.2d_all}
\end{figure}
Also in this case, as in the three-dimensional one, the spin-orbit produces
interesting effects in the condensate fraction, showing a nodal
point at $\epsilon_B\approx \epsilon_F$, set in when $v_R\simeq v_F$,
see Fig.~\ref{fig.2d_f1} and Fig.~\ref{fig.2d_f2}, in the neighborhood of
which the slope of the curve decreases. As a result, the spin-orbit coupling
promotes the singlet condensation in the BCS side and suppresses it on the BEC side.
We observe that, contrary to the 3D case, in 2D the condensate fraction
approaches the value of 1 only for an extremely large interaction strength,
i.e. for $\epsilon_B/\epsilon_F \gg 1$.
Again, the chemical potential $\mu$ is pushed toward more negative values 
(see the upper panel of Fig.~\ref{fig.2d_all}).
For small spin-orbit coupling, and $\epsilon_B\rightarrow 0$, one recovers
the known perturbative result, $\mu\simeq \epsilon_F- mv_R^2/2$.
The pairing gap $\Delta$ (lower panel of Fig.~\ref{fig.2d_all}) 
is, instead, increased by the Rashba
spin-orbit interaction in the whole crossover.

\section{Conclusions}

We have studied the evolution of BCS superconductors to BEC superfluids in
the presence of an artificial spin-orbit coupling of Rashba and/or Dresselhaus
type in two and three dimensions. We have shown that, unlike the chemical
potential and the pairing gap which exhibit no particular
behaviors at the crossover, the condensate fraction is very peculiar.
The condensation of singlet pairs, in fact, is promoted by Rashba 
coupling in the BCS
regime whereas it is suppressed in the BEC regime. In the middle,
both in three and in two dimensions and for large enough
Rashba spin-orbit coupling, there is a nodal point where the curves of the
singlet condensate fraction cross each other, and, for this reason, this 
can be considered the putative point of the crossover.
On the other hand, the triplet contribution to the condensate fraction has not a monotonic behavior as a function of the scattering parameter, swelling up close to the crossover. 
The full condensate fraction increases with the spin-orbit interaction. 
Because in our calculations we have used the mean-field theory,
it is important to stress that Monte Carlo simulations
have shown that, at zero temperature, beyond-mean-field effects
are negligible in the BCS side of the BCS-BEC crossover, whereas they become
relevant in the deep BEC side \cite{astrakharchik,bertaina}.
In conclusion, we think that our results can be of interest
for future experiments with artificial gauge potentials
in degenerate gases made of alkali-metal atoms.

\begin{acknowledgments}
L.D. thanks the International School for Advanced Studies, SISSA,
Trieste, for hospitality and facilities during the completion of this work. 
\end{acknowledgments}

\end{document}